\documentclass{article}
\usepackage[T1]{fontenc} 
\usepackage[utf8]{inputenc} 
\usepackage{ismir,amsmath,cite,url}
\usepackage{graphicx}
\usepackage{color}
\usepackage{mathrsfs}
\usepackage{amsfonts}
\usepackage{enumitem}
\usepackage{titlesec}

\usepackage[ruled,vlined]{algorithm2e}
 
\SetCommentSty{mycommfont}
\usepackage{subcaption}

\usepackage[bookmarks=false,hidelinks]{hyperref}
\newcommand{\code}[1]{{\small{\texttt{#1}}}}
\newcommand{\score}{\code{score}}
\newcommand{\simunote}{\code{simu\_note}}
\newcommand{\note}{\code{note}}
\newcommand{\n}{\code{n}}
\newcommand{\sn}{\code{sn}}
\newcommand{\scr}{\code{sc}}

\titlespacing*{\section}{0pt}{2ex}{1ex}
\titlespacing*{\subsection}{0pt}{1.5ex}{0.5ex}
\titlespacing*{\subsubsection}{0pt}{1ex}{0.5ex}
\title{PIANOTREE VAE: Structured Representation Learning for Polyphonic Music}

\multauthor
{Ziyu Wang$^1$ \hspace{1cm} Yiyi Zhang$^2$ \hspace{1cm} Yixiao Zhang$^1$ \hspace{1cm} Junyan Jiang$^1$} { \bfseries{\hspace{1cm} Ruihan Yang$^1$ \hspace{1cm} Junbo Zhao (Jake)$^3$ \hspace{1cm} Gus Xia$^1$}\\
 $^1$ Music X Lab, Computer Science Department, NYU Shanghai\\
$^2$ Center for Data Science, New York University\\
$^3$ Computer Science Department, Zhejiang University\\
{\tt\small \{ziyu.wang, yz2092, yixiao.zhang, jj2731, ry649, j.zhao, gxia\}@nyu.edu}
}
\def\authorname{Ziyu Wang, Yiyi Zhang, Yixiao Zhang, Junyan Jiang, Ruihan Yang, Junbo Zhao (Jake), Gus Xia}

\begin{document}

\maketitle
\begin{abstract}
The dominant approach for music representation learning involves the deep unsupervised model family \emph{variational autoencoder} (VAE).
However, most, if not all, viable attempts on this problem have largely been limited to monophonic music.
Normally composed of richer modality and more complex musical structures, the polyphonic counterpart has yet to be addressed in the context of music representation learning.
In this work, we propose the PianoTree VAE, a novel tree-structure extension upon VAE aiming to fit the polyphonic music learning. 
The experiments prove the validity of the PianoTree VAE via (i)-semantically meaningful latent code for polyphonic segments; (ii)-more satisfiable reconstruction aside of decent geometry learned in the latent space; (iii)-this model’s benefits to the variety of the downstream music generation.\!\!\footnote{Code and demos can be accessed via \url{https://github.com/ZZWaang/PianoTree-VAE}}

\end{abstract}
\section{Introduction}\label{sec:introduction}

Unsupervised machine learning has led to a marriage of symbolic learning and vectorized representation learning \cite{nlp-vae1, nlp-vae2, recurrent-vae}. In the computer music community, the MusicVAE \cite{musicvae} enables the interpolation in the learned latent space to render some smooth music transition. The EC$^2$-VAE \cite{ec2vae} manages to disentangle certain interpretable factors in music and also provides a manipulable generation pathway based on these factors. Pati \textit{et al}. \cite{lerchvae} further utilizes the recurrent networks to learned music representations for longer-term coherence. 


Unfortunately, most of the success has been limited to monophonic music. 
The generalization of the learning frameworks to polyphonic music is not trivial, due to its much higher dimensionality and more complicated musical syntax. 
The commonly-adopted MIDI-like event sequence modeling or the piano-roll formats fed to either recurrent or convolutional networks have fell short in learning good representation, which usually leads to unsatisfied generation results \cite{musegan, yang2017midinet, mt-vae}. In this paper, we hope to pioneer the development of this challenging task.
To begin with, we conjecture a proper set of \textbf{inductive bias} for the desired framework: (i)-a sparse encoding of music as the model input; (ii)-a neural architecture that incorporates the hierarchical structure of polyphonic music (i.e., musical syntax).

\begin{figure}[t]
 \centerline{
 \includegraphics[width=0.8\columnwidth]{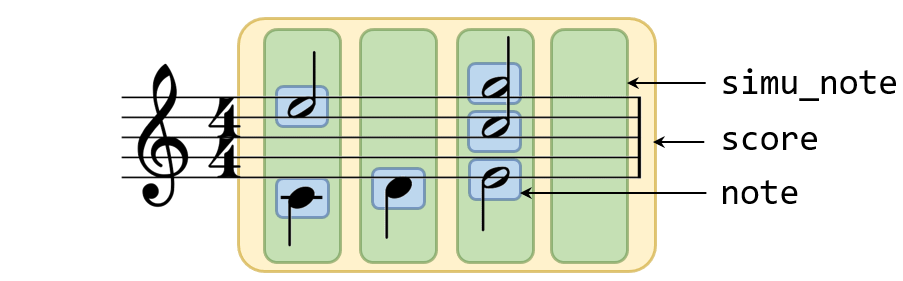}}
 \caption{An illustration of the proposed polyphonic syntax.}
 \label{fig:example}
\end{figure}

Guided by the aforementioned design principles, we propose PianoTree VAE, a hierarchical representation learning model under the VAE framework. We adopt a tree structured musical syntax that reflects the hierarchy of musical concepts, which is shown in \figref{fig:example}. In a top-down order: we define a \score{} (indicated by the yellow rectangle) as a series of \simunote{} events (indicated by the green rectangles), a \simunote{} as multiple \note{} events sharing the same onset (indicated by blue rectangles), and each \note{} has several attributes such as \textit{pitch} and \textit{duration}. In this paper, we focus on a simple yet common form of polyphonic music---piano score, in which each note has only pitch and duration attributes. For future work, this syntax can be generalized to multiple instruments and expressive performance by adding extra attributes such as voice, expressive timing, dynamics, etc.


The whole neural architecture of PianoTree VAE can be seen as a tree. Each node represents the embedding of either a \score{}, \simunote{}, or \note{}, where a higher level representation has larger receptive fields. The edges are bidirectional where a recurrent module is applied to either encode the children into the parent or decode the parent to generate its children.


Through  extensive  evaluations, we show that PianoTree VAE yields semantically more meaningful latent representations and further downstream generation quality gains, on top of the current state-of-the-art solutions.


\section{Related Work}\label{sec:2}

The complex hierarchical nature of music data has been studied for nearly a century (e.g. GTTM \cite{gttm}, Schenklerian Analysis \cite{schenkerian}, and their follow-up works \cite{gttmresearch1, gttmresearch2, schenresearch1, schenresearch2}). However, the emerging deep representation-learning models still lack the compatible solutions to deal with the complex musical structure. In this section, we first review different types of polyphonic music generation in Section~\ref{sec:2:polytype}. After that, we discuss some popular deep music generative models indexed by their compatible data structure from Section~\ref{sec:2:pr} to Section~\ref{sec:2:gnn}.

%

\subsection{Different Types of Polyphony}\label{sec:2:polytype}
In the context of deep music generation, \textit{polyphony} can refer to three types of music: 1) multiple monophonic parts (e.g., a four-part chorus), 2) a single part of a polyphonic instrument (e.g., a piano sonata), and 3) multiple parts of polyphonic instruments (e.g., a symphony). 

The first type of polyphonic music can be created by simply extending the number of voices in monophonic music generation with some inter-voice constraints. Some representative systems belonging to this category include DeepBach \cite{deepbach}, XiaoIce \cite{xiaoice}, and Coconet \cite{huang2017counterpoint}. Music Transformer \cite{music-transformer} and the proposed PianoTree VAE both focus on the generation of the second type of polyphony, which is a much more difficult task. Polyphonic pieces under the second definition no longer have a fixed number of ``voices'' and consist of more complex textures. The third type of polyphony can be regarded as an extension of the second type, and we leave it for future work.

\subsection{Piano-roll and Compatible Models}\label{sec:2:pr}
Piano-roll and its variations \cite{jambot, deepj, musegan, rethinking} view polyphonic music as 3-D (one-hot) tensors, in which the first two dimensions denote time and pitch and the third dimension indicates whether the token is an onset, sustain or rest.
A common way for deep learning models to encode/decode a piano-roll is to use recurrent layers along the time-axis while the pitch-axis relations are modeled in various ways \cite{jambot, deepj,  boulangerg}. Another method is to regard a piano-roll as an image with three channels (onset, sustain and rest) and apply convolutional layers \cite{musegan, rethinking}. 

Through the proposal of PianoTree VAE, we argue that a major way to improve the current deep learning models is to utilize the built-in priors (intrinsic structure) in the musical data. In our work, we primarily use the sparsity and the hierarchical priors.


\subsection{MIDI-like Event Sequence and Compatible Models}\label{sec:2:midimsg}
MIDI-like event sequence is first used in deep music generation in performanceRNN \cite{perf-rnn} and Multi-track MusicVAE \cite{mt-vae}, and then broadly applied in transformer-based generation \cite{music-transformer, lakhnes, poptfm}. 
This direction of work leverages the sparsity of polyphonic data to efficiently flatten polyphonic music into an array of events. The vocabulary size of events usually tripples the vocabulary size of MIDI pitches, including ``note-on'' and ``note-off'' events for 128 MIDI pitches, ``time shifts'', and so on. 

However, the format of MIDI-like events lacks the proper flexibility. A few operations are made difficult due to its very nature. For instance, during addition or deletion of notes, often numerous ``time shift'' tokens must be merged or split with the ``note-on'' or ``note-off'' tokens being changed all-together. This has caused the model being trained inefficient for the potential generation tasks. In addition, this format has a risk of generating illegal sequences, say a ``note on'' message without a paired ``note off'' message.

Similarly, we see the note-based approaches \cite{crnngan, tfm-nade}, in which polyphonic music is represented as a sequence of note tuples, as an alternative to the MIDI-like methods. The representation has resolved the illegal generation problem but still not revealed much of the intrinsic music structure. We argue that our work improves on the note-based approaches by utilizing deeper musical structures implied by the data. (See Section 3.1 for details.)


\subsection{GNN as a Novel Structure}\label{sec:2:gnn}
Recently, we see a trend in using graph neural networks (GNN) \cite{gnn} to represent polyphonic score \cite{musicgnn}, in which each vertex represents a note and the edges represent different musical relations.
Although the GNN-based model offers sparse representation learning capacity, it is limited by the specification of the graph structure design and it is nontrivial to generalize it for score generations.


\section{Method}\label{sec:3}

\subsection{Data Structure}\label{sec:3:datastruct}

We first define a data structure to represent a polyphonic music segment, which contains two components: 1) \textit{surface structure}, a data format to represent the music observation, and 2) \textit{deep structure}, a tree structure (containing \score{}, \simunote{} and \note{} nodes) showing the syntactic construct of the music segment. 

Each music segment lasts $T$ time steps with $\frac{1}{4}$ beat as the shortest unit. We further use $K_t$, where $1 \leq t \leq T$ to denote the number of notes having the same onset $t$. The current model uses $T=32$, i.e., each music segment is 8-beat long.

\subsubsection{Surface Structure}
The surface structure is a nested array of \textit{pitch-duration} tuples, denoted by $\{(p_{t, k}, d_{t, k})|1\leq t \leq T, 1 \leq k \leq K_t\}$. Here, $(p_{t, k}, d_{t, k})$ is the $k^{\text{th}}$ lowest note starting at time step $t$. The pitch attribute $p_{t, k}$ is a 128-D one-hot vector corresponding to 128 MIDI pitches. The duration attribute $d_{t, k}$ encodes the duration ranging from 1 to $T$ using a $\log_2T$-bit binary vector. For example, when $T=32$ ($\log_2T = 5$), `00000’ represents a $16^{\text{th}}$ note, ‘00001’ is an $8^{\text{th}}$ note, ‘00010’ is a dotted $8^{\text{th}}$ note, and so on so forth. The base-2 design is inspired by the similar binary relation among different note values in western musical notation. 

The bottom part of \figref{fig:data_embedding} illustrates the surface structure of the music example in \figref{fig:example}. We see that the data structure is a sparse encoding of music, and it eliminates illegal tokens since every possible nested array has a correspondent music.

\subsubsection{Deep Structure}
\begin{figure}[tb]
 \centerline{
 \includegraphics[width=\columnwidth]{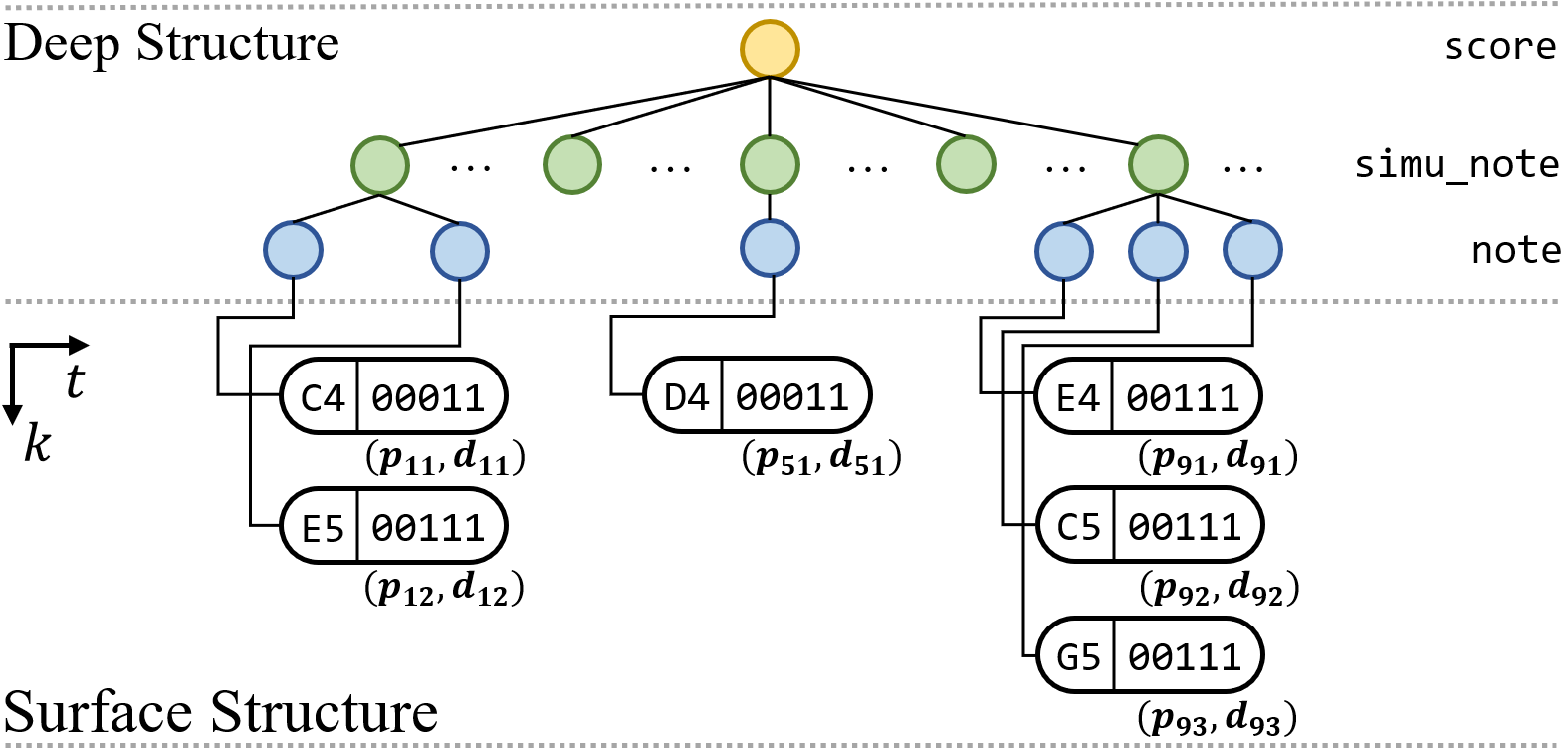}}
 \caption{An illustration of PianoTree data structure to encode the music example in \figref{fig:example}.}
 \label{fig:data_embedding}
\end{figure}

We further build a syntax tree to reveal the hierarchical relation of the observation. First, for $1\leq t \leq T, 1 \leq k \leq K_t$, we define $\text{\note{}}_{t, k}$ as the summary (i.e., embedding) of $(p_{t, k}, d_{t, k})$, which are the bottom layers of the tree. Then, for $1\leq t \leq T$, we define $\text{\simunote{}}_{t}$ as the summary of $\text{\note{}}_{t, 1 \leq k \leq K_t}$, which are the middle layers of the tree. Finally, we define the $\text{\score{}}$ as the summary of $\text{\simunote{}}_{1 \leq t \leq T}$, which is the root of the tree. The upper part of \figref{fig:data_embedding} illustrates the deep structure built upon its surface structure.

The syntax tree, so-called the deep structure has both musical and linguistic consideration. In terms of music, \note{}, \simunote{} and \score{} roughly reflect the musical concept of a note, chord and grouping. In terms of linguistics, the tree is analogous to a constituency tree, with surface structure being the terminal nodes and deep structure being the non-terminals. Recent studies in natural language processing have revealed that incorporating natural language syntax results in better semantics modeling \cite{syntax1, syntax2}.

 \begin{figure}[tb]
 \centerline{
 \includegraphics[width=\columnwidth]{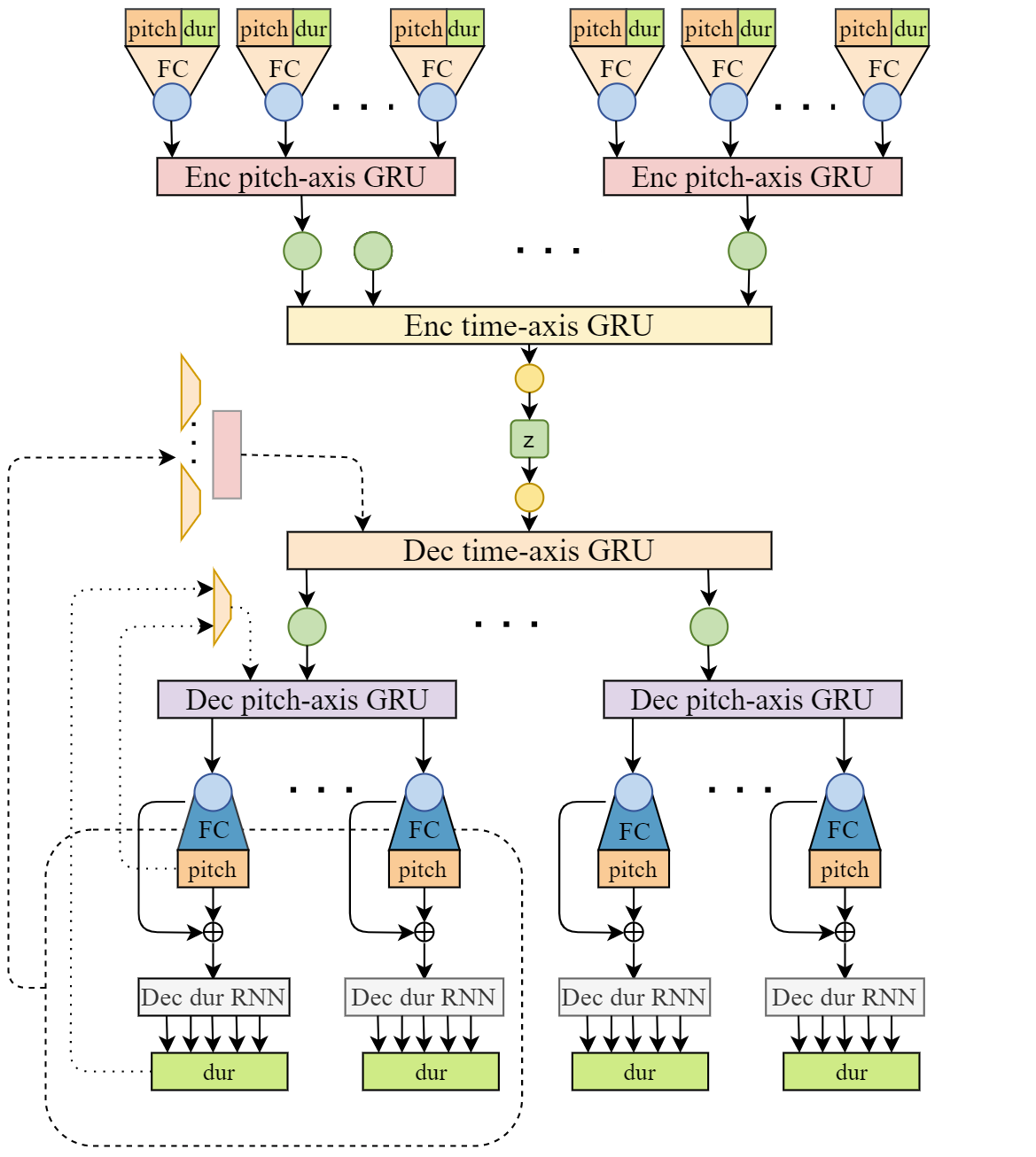}}
 \caption{An overview of the model architecture. The recurrent layers are represented by rectangles and the fully-connected (FC) layers are represented by trapezoids. The \note{}, \simunote{} and \score{} events are represented by circles.}
 \label{fig:model}
\end{figure}


\subsection{Model Structure}\label{sec:3:modelstruct}
We use the surface structure of polyphonic music as the model input. The VAE architecture is built upon the deep structure.


We denote the music segment in the proposed surface structure as $x$ and the latent code as $z$, which conforms to a standard Gaussian prior denoted by $p(z)$. The encoder models the approximated posterior $q_\phi(z|x)$ in a bottom-up order of the deep structure. First, \note{} embeddings are computed through a linear transform of pitch-duration tuples. Second, the \note{} embeddings (sorted by pitch) are then embedded into \simunote{} using a bi-directional GRU \cite{gru} by concatenating the last hidden states on both ends. With the same method, the \simunote{} embeddings (sorted by onsets) are summarized into \score{} by another bi-directional GRU. We assume an isotropic Gaussian posterior, whose mean and log standard deviation are computed by a linear mapping of \score{}. Algorithm~{\ref{algo:enc}} shows the details.


\begin{algorithm}[]
\SetAlgoLined
\SetKwInOut{Input}{input}
\tcc{gru($\cdot$): passes a sequence to bi-directional GRU and ouputs the concatenation of hidden states from both ends.}
 \Input{PianoTree $x = \{(p_{t, k}, d_{t, k}), 1\leq t \leq T, 1\leq k \leq K_t$\}}
 \lForEach{t, k}{\n{}$_{t, k}$ $\gets$ emb$_{\text{enc}}(p_{t, k}, d_{t, k})$}
 \lForEach{t}{\sn$_{t}$ $\gets$ gru$_{\text{enc}}^{\text{pitch}}$(\n{}$_{t, 1:K_t}$)}
 \scr{} $\gets$ gru$_{\text{enc}}^{\text{time}}$(\sn{}$_{1: T}$) \;
 $\mu \gets \text{fc}_\mu$(\scr{}); $\sigma \gets \exp(\text{fc}_\sigma$(\scr{})) \;
 \Return $q(z|x) = N(\mu, \sigma^2)$ \;
 \caption{The \textbf{PianoTree Encoder}. \n{}, \sn{}, \scr{} are short for \note{}, \simunote{}, \score{}.}
 \label{algo:enc}
\end{algorithm}

The decoder models $p_\theta(x|z)$ in a top-down order of the deep structure, almost mirroring the encoding process. We use a uni-directional time-axis GRU to decode \simunote{}, another uni-directional (pitch-axis) GRU to decode \note{}, a fully connected layer to decode pitch attributes, and finally another GRU to decode duration attribute starting from the most significant bit. Algorithm~\ref{algo:dec} shows the details.

We use the ELBO (evidence lower bound) \cite{vae} as our training objective. Formally,

\begin{equation}
\mathcal{L}(\phi, \theta; x) = -\mathbb{E}_{z\sim q_\phi}
\log p_\theta(x|z)  + \beta \mathrm{KL}\Bigl(q_\phi||p(z)\Bigr)\text{,}
\end{equation}
where $\beta$ is a balancing parameter used in $\beta$-VAE \cite{betavae}.


We denote the embedding size of \note{}, \simunote{} and \score{} as $e_\text{n}$, $e_{\text{sn}}$ and $e_{\text{sc}}$; the dimension of latent space as $d_\text{z}$; and the hidden dimensions or pitch-axis, time-axis and dur GRUs as $h_\text{p}$, $h_\text{t}$ and $h_\text{d}$ respectively. In this work, we report our result on the following model size: $e_\text{n}=128$, $e_\text{sn} = h_{\text{p}, \text{dec}} = 2 \times h_{\text{p}, \text{enc}}=512$, $e_{\text{sc}}=h_{\text{t}, \text{dec}} = 2 \times h_{\text{t}, \text{enc}}=1024$, $h_{\text{d}, \text{dec}}=64$ and $d_z=512$. 

\section{Experiments}\label{sec:4}
In this section, we compare PianoTree VAE with several baseline models. We present the dataset in Section~\ref{sec:4:dataset}, baseline models in Section~\ref{sec:4:baseline},and the training details in Section~\ref{sec:4:training}. We present the objective evaluation on reconstruction accuracy in Section~\ref{sec:4:objeval}. In Section~\ref{sec:4:visualize}, we inspect and visualize the latent space of \note{} and \simunote{}. After that, we present the subjective evaluation on latent space traversal in Section~\ref{sec:4:subeval}.  Finally, we apply the learned representation to downstream music generation task in Section~\ref{sec:4:musicgen}.


\begin{algorithm}[t]
\SetAlgoLined
\SetKwInOut{Input}{input}
\tcc{gru($\cdot$), same as Algorithm 1. \\
grucell($\cdot$, \!\!\!$\cdot$): updates the hidden state using the current input and the previous hidden state. The output is replicated.}
 \Input{latent representation $z$}
 \scr{}  $\gets z$ \;
 $\tilde{\text{\sn{}}}_{0}$, $\tilde{\text{\n{}}}_{:, 0}$, $d_{:, :, 0} =$ \textup{<SOS>}\;
 \For{$t = 1, 2, ... T$}{
  [\sn{}$_t$, \scr{}] $\gets$ grucell$_{\text{dec}}^{\text{time}}$($\tilde{\text{\sn{}}}_{t - 1}$, \scr{})\;
  \For{$k = 1, 2, ...$}{
  [\n{}$_{t, k}$, \sn{}$_{t}$] $\gets$ grucell$_{\text{dec}}^{\text{pitch}}$($\tilde{\text{\n{}}}_{t, k-1}$, \sn{}$_{t}$) \;
  $p_{t, k}$ $\gets$ softmax(fc(\n{}$_{t, k}$)) \;
  \For{$r = 1, 2, ..., 5$}{
  $h$ = [\n{}$_{t, k}$, $p_{t, k}$] \;
  [$y_{t, k, r}$, $h$] = grucell$_{\text{dec}}^{\text{dur}}$($d_{t, k, r-1}$, $h$)\;
  $d_{t, k, r}$ $\gets$ softmax($y_{t, k, r}$)\;
  }
  $d_{t, k} = [d_{t, k, 1:5}]$ \;
  \lIf{$p_{t, k}$ $\neq$ \textup{<EOS>}}{$K_t \gets k$; break}
  $\tilde{\text{\n{}}}_{t, k}$ $\gets$ emb$_{\text{enc}}$($p_{t, k}$, $d_{t, k}$) \;
  }
  $\tilde{\text{\sn{}}}_t$ $\gets$ gru$_{\text{enc}}^{\text{pitch}}$(\n{}$_{t, 1:K_t}$)\;
 }
 \Return $\{(p_{t, k}, d_{t, k}), 1\leq t \leq T, 1\leq k \leq K_t\}$ \;
 \caption{The \textbf{PianoTree Decoder}. We still use the abbreviation \n{}, \sn{}, and \scr{}, defined in Algorithm~\ref{algo:enc}}
 \label{algo:dec}
\end{algorithm}


\subsection{Dataset}\label{sec:4:dataset}
We collect around 5K classical and popular piano pieces from Musicalion\footnote{Musicalion: \url{https://www.musicalion.com}.} and the POP909 dataset \cite{pop909}. We only keep the pieces with $\frac{2}{4}$ and $\frac{4}{4}$ meters and cut them into 8-beat music segments (i.e., each data sample in our experiment contains 32 time steps under sixteenth note resolution). In all, we have 19.8K samples. We randomly split the dataset (at song-level) into training set (90\%) and test set (10\%). All training samples are further augmented by transposing to all 12 keys.

\subsection{Baseline Model Architectures}\label{sec:4:baseline}

We train four types of baseline models in total using piano-roll (Section~\ref{sec:2:pr}) and MIDI-like events (Section~\ref{sec:2:midimsg}) data structures. As a piano-roll can be regarded as either a sequence or a 2-dimensional image, we couple it with three neural encoder-decoder architectures: a recurrent VAE (\textbf{pr-rnn}), a convolutional VAE (\textbf{pr-cnn}), and a fully-connected VAE (\textbf{pr-fc}). For the MIDI-like events, we couple it with a recurrent VAE model (\textbf{midi-seq}). All models share the same latent space dimension ($d_z=512$). Specifically,

\begin{itemize}[leftmargin=*, topsep=0pt,parsep=0pt,itemsep=0pt,partopsep=0pt]
    \item The piano-roll recurrent VAE (\textbf{pr-rnn}) model is similar to a 2-bar MusicVAE proposed in \cite{musicvae}. The hidden dimensions of the GRU encoder and decoder are both 1024.  
    \item The piano-roll convolutional VAE (\textbf{pr-cnn}) architecture adopts a convolutional--deconvolutional architecture. The encoder contains 8 convolutional layers with kernel size $3\times3$. Strided convolution is performed at the 3$^\text{rd}$, 5$^\text{th}$, 7$^\text{th}$ and 8$^\text{th}$ layer with stride size $(2\times1),(2\times3),(2\times2)$ and $(2\times2)$ respectively. The decoder adopts the deconvolution operations in a reversed order.
    \item The piano-roll fully-connected VAE (\textbf{pr-fc}) architecture uses a time-distributed 256-dimensional embedding layer, followed by 3 fully-connected layers with the hidden dimensions [1024, 768] for the encoder. The decoder adopts the counter-operations in a reversed order.  
    \item The MIDI-like event recurrent VAE (\textbf{midi-seq}) adopts the recurrent model structure similar to \textbf{pr-rnn}. Here, the event vocabulary contains 128 ``note-on'', 128 ``note-off'' and 32 ``time shift'' tokens. The embedding size of a single MIDI event is 128. The hidden dimensions of the encoder GRU and decoder GRU are 512 and 1024 respectively.

\end{itemize}

\subsection{Training}\label{sec:4:training}
For all models, we set batch size $=128$ and use Adam optimizer \cite{adam} with a learning rate starting from 1e-3 with exponential decay to 1e-5. For PianoTree VAE, we use teacher forcing \cite{teacherforcing} for decoder time-axis and pitch-axis GRU and for other recurrent-based baselines, we use teacher forcing in the decoders. The teacher forcing rates start from 0.8 and decay to 0.0. PianoTree VAE converges within 6 epochs, and the baseline models converge in approximately 40 to 60 epochs. 

\begin{table}[ht]
\centering
\resizebox{\columnwidth}{!}{
\begin{tabular}{lccccc}
 \hline
 \small{Models} &\small{\textbf{PianoTree}} & \small{\textbf{midi-seq}}  & \small{\textbf{pr-rnn}} & \small{\textbf{pr-cnn}} & \small{\textbf{pr-fc}}  \\ \hline
 
 \small{Onset Precision} & \small{\textbf{0.9558}}  & \small{0.8929} & \small{0.9533} & \small{0.9386} & \small{0.9211}  \\ 
 
  \small{Onset Recall} & \small{\textbf{0.9532}}  & \small{0.6883} & \small{0.9270} & \small{0.8818} & \small{0.8827}\\
 
 \hline
 
 \small{Onset F1} & \small{\textbf{0.9545}}  & \small{0.7774} & \small{0.9399} & \small{0.9093} & \small{0.9015}  \\
  
 \hline\hline
 
 \small{Duration Precision} & \small{\textbf{0.9908}}  & \small{0.3826} & \small{0.9777} & \small{0.9757} & \small{0.9688}  \\ 
 
  \small{Duration Recall} & \small{0.9830}  & \small{\textbf{0.9899}} & \small{0.9891} & \small{0.9796} & \small{0.9743}\\
 
 \hline
 
 \small{Duration F1} & \small{\textbf{0.9869}}  & \small{0.5519} & \small{0.9834} & \small{0.9777} & \small{0.9715}  \\
 
 \hline
\end{tabular}
}
\caption{Objective evaluation results on reconstruction criteria. PianoTree is our proposed method. Other columns correspond to the baseline models described in Section~\ref{sec:4:baseline}.}
\label{tbl:result}
\end{table}


\begin{figure*}[t]
 \centerline{
 \includegraphics[width=\textwidth]{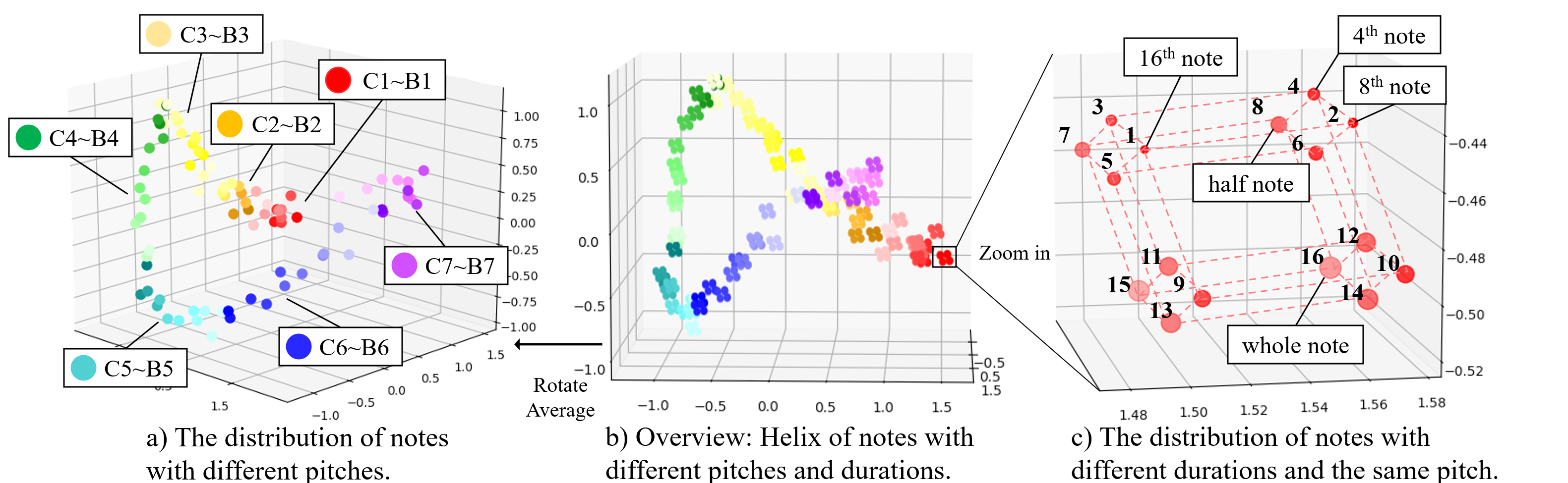}}
 \caption{A visualization of \note{} embeddings after dimensionality reduction using PCA.}
 \label{fig:note}
\end{figure*}

\subsection{Objective Evaluation of Reconstruction}\label{sec:4:objeval}
The objective evaluation is performed by comparing different models in terms of their reconstruction accuracy of pitch onsets and note duration\cite{metric, mir_eval}, which are commonly used measurements in music information retrieval tasks. For note duration accuracy, we only consider the notes whose onset and pitch reconstruction is correct. Table~\ref{tbl:result} summarizes the results where we see that the PianoTree 
VAE (the 1$^{\text{st}}$ column) is better than others in terms of F1 score for both criteria.

\subsection{Latent Space Visualization}\label{sec:4:visualize}

\figref{fig:note} shows the \textit{latent note space} by plotting different \note{} embeddings after dimensionality reduction by PCA (with the three largest principal components being reserved). Each colored dot is a \note{} embedding and a total of 1344 samples are displayed; note pitch ranges from C-1 to C-8 and note duration from a sixteenth note to a whole note.

We see that the \note{} embeddings have the desired geometric properties. \figref{fig:note} (a) \& (b) show that at a macro level, notes with different pitches are well sorted and form a ``helix'' in the 3-D space. \figref{fig:note} (c) further shows that at a micro level, 16 different note durations (with the same pitch) form a ``fractal parallelogram'' due to the binary encoding of duration attributes. One of the advantages of the encoding method is the translation invariance property. For example, the duration difference between the upper left cluster and the lower left cluster is 8 semiquavers, so is the difference between the upper right and lower right cluster. The same property also applies to the four smaller-scale parallelograms. 


\begin{figure}[htb]
 \centerline{
 \includegraphics[width=\columnwidth]{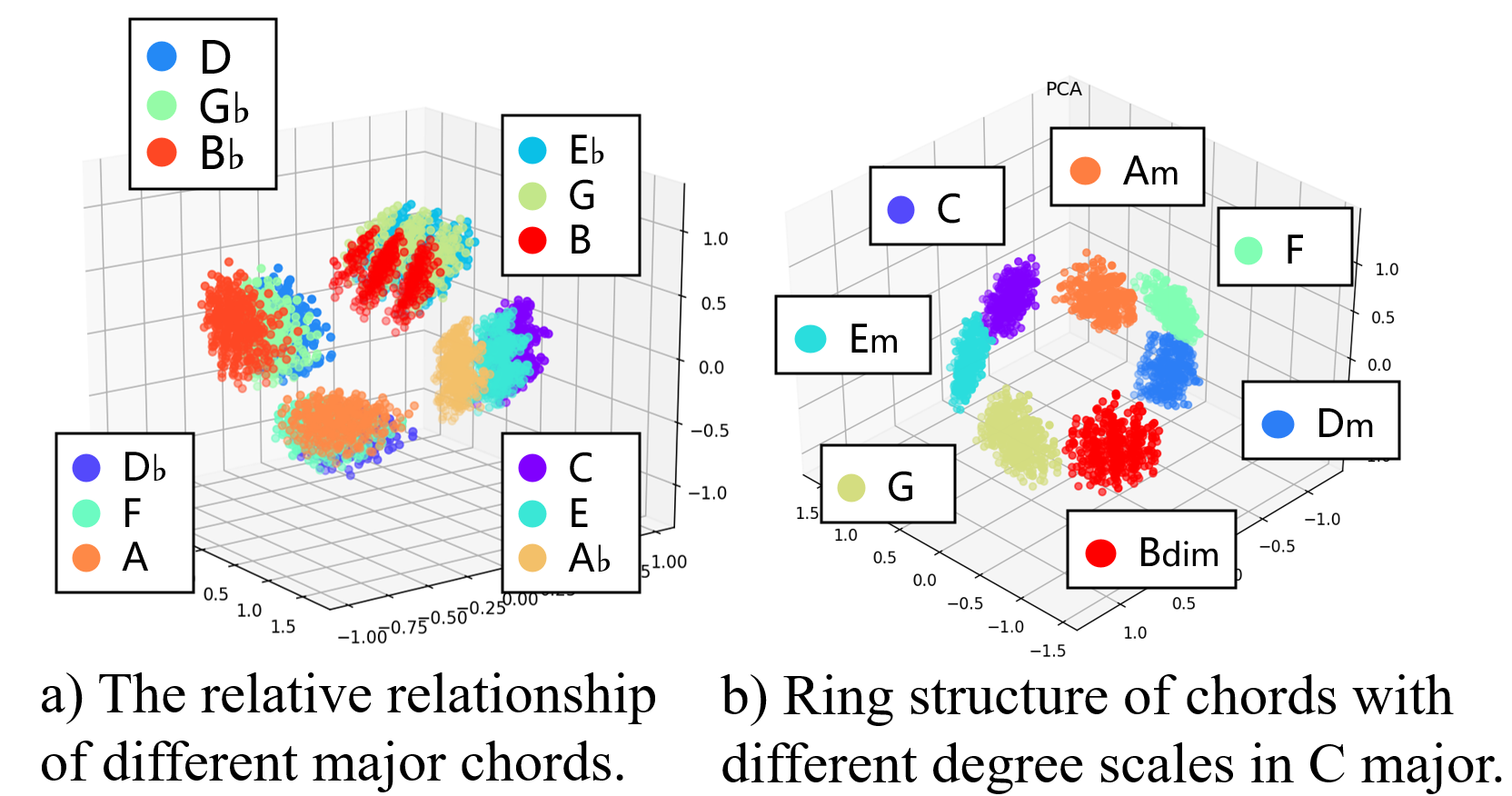}}
 \caption{A visualization of \simunote{} embeddings after dimensionality reduction using PCA.}
 \label{fig:simu_note}
\end{figure}

\figref{fig:simu_note} is a visualization of the latent chord space by plotting different \simunote{} embeddings under PCA dimensionality reduction. Each colored cluster corresponds to a chord label realized in 343 different ways (we consider all possible pitch combinations within three octaves, with a minimum of 3 notes and a maximum of 9 notes). The duration for all chords is one beat.

The geometric relationships among different chords are consistent and human interpretable. In specific, \figref{fig:simu_note} (a) shows the distribution of 12 different major chords, which are clustered in four different groups. By unfolding the circle in a counterclockwise direction, we can observe the existence of \textit{the circle of the fifth}. \figref{fig:simu_note} (b) is the visualization of seven C major triad chords: forming a ring in the order of 1-3-5-7-2-4-6 degree in the counterclockwise direction.

\subsection{Subjective Evaluation of Latent Space Interpolation}\label{sec:4:subeval}
Latent space traversal \cite{musicvae, our-nime-vae, ec2vae} is a popular technique to demonstrate model generalization and the smoothness of the learned latent manifold. When interpolating from one music piece to another in the latent space, new pieces can be generated by mapping the representations back to the signals. If a VAE is well trained, the generated piece will sound natural and form a smooth transition.

To this end, we invite people to subjectively rate the models through a double-blind online survey. During the survey, the subjects first listen to a pair of music, and then listen to 5 versions of interpolation, each generated by a model listed in Table~\ref{tbl:result}. Each version is a randomly selected pair of music segments, and the interpolation is achieved using SLERP\cite{slerp}. Since the experiment requires careful listening and a long survey could decrease the quality of answers, each subject is asked to rate only 3 pairs of music, i.e., $3 \times 5 = 15$ interpolations in a random order. After listening to the 5 interpolations of each pair, subjects are asked to select two best versions: one in terms of the \textit{overall musicality}, and the other in terms of the \textit{smoothness of transition}.

\begin{figure}[htb]
 \centerline{
 \includegraphics[width=\columnwidth]{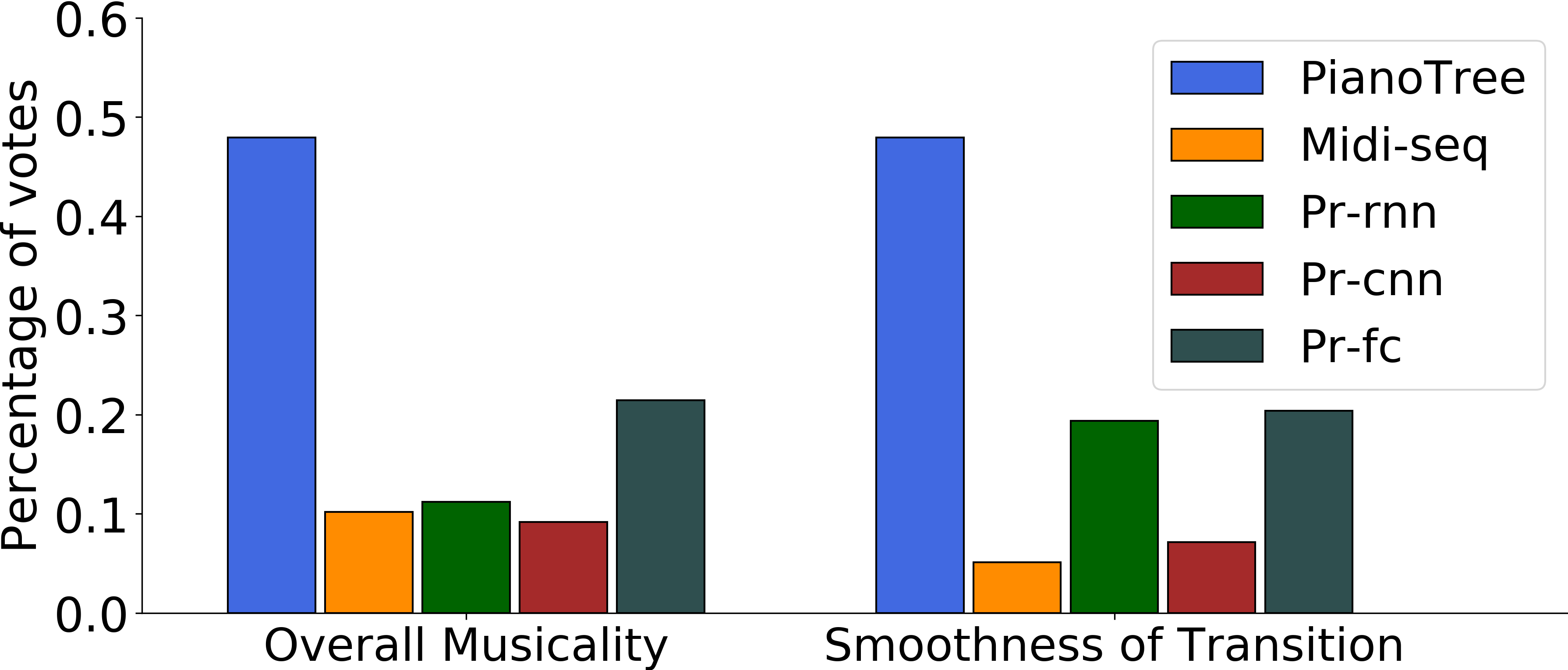}}
 \caption{Subjective evaluation results of latent space interpolation.}
 \label{fig:sub-exp}
  \vspace{-1\baselineskip}
\end{figure}

A total of n = 33 subjects (12 females and 21 males) with different music backgrounds have completed the survey. The aggregated result (as in \figref{fig:sub-exp}) shows that the interpolations generated by our model are better than the ones generated by baselines, in terms of both overall musicality and smoothness of transition. Here, different colors represent different models (with the blue bars being our model and other colors being the baselines), and the height of the bars represent the percentage of votes (on the best candidate).

\subsection{Downstream Music Generation}\label{sec:4:musicgen}
In this section, we further explore whether the polyphonic representation helps with \textit{long-term music generation} when coupled with standard downstream sequence prediction models. (Similar tasks have been applied to \textit{monophonic music} in \cite{ke} and \cite{lerchvae}.)

The generation task is designed in the following way: given 4 measures of piano composition, we predict the next 4 measures using a Transformer decoder (as in \cite{tfm}). We compare three different music representations: MIDI-like event sequence (Section~\ref{sec:2:polytype}), pretrained (decoder) \simunote{} embeddings, and latent vector \textit{z} for every 2-measure music segment (without overlap). Here \textit{z} is the mean of the approximated posterior from the encoder. For all three representations, we use the same Transformer decoder architecture (outputs of dimension = 128, number of layers = 6 and number of heads = 8) with the same training procedure. Only the loss functions are correspondingly adjusted based on different representations: cross entropy loss is applied to midi-event tokens and MSE loss is applied to both \simunote{} and latent vector \textit{z}. We use the same datasets mentioned in Section~\ref{sec:4:dataset} and cut the original piano pieces into 8-measure subsequent clips for generation purposes. We still keep 90\% for training and 10\% for testing.

We then invited people to subjectively rate different music generations through a double-blind online survey (similar to the one in Section~\ref{sec:4:subeval}). Subjects are asked to listen to and rate 6 music clips, each of which contains 3 versions of 8-measure generation using different note representations. Subjects are told that the first 4 measures are given and the rest are generated by the machine. For each music clip, subjects rate it based on \textit{creativity}, \textit{naturalness} and \textit{musicality}. 

\begin{figure}[htb]
 \centerline{
 \includegraphics[width=\columnwidth]{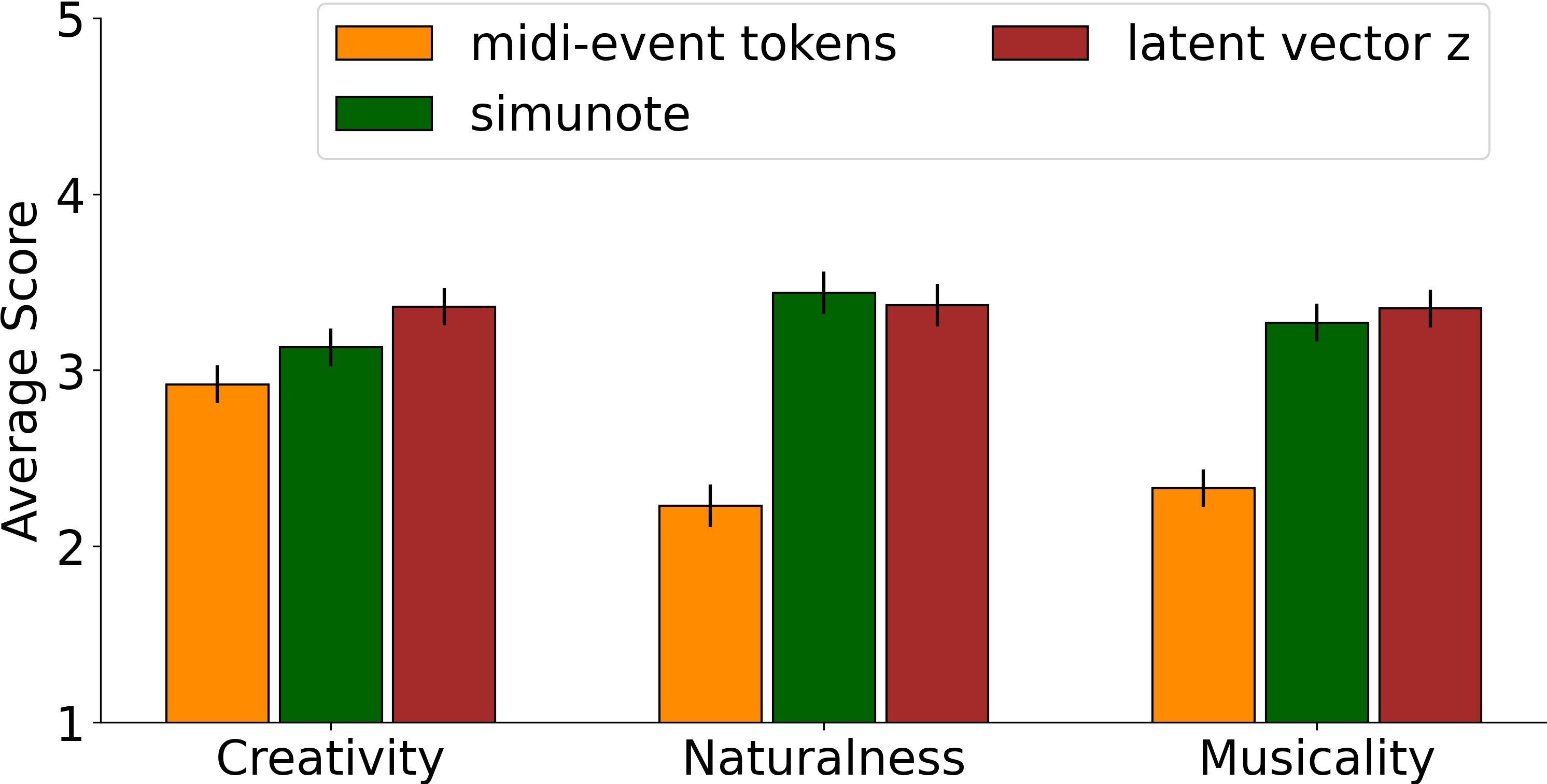}}
 \caption{Subjective evaluation results of downstream music generation.}
 \label{fig:generation}
 
\end{figure}

A total of n = 48 subjects (20 females and 28 males) with different music backgrounds have participated in the survey. \figref{fig:generation} summarizes the survey results, where
the heights of bars represent means of the ratings and the error bars represent the confidence intervals computed via within-subject
ANOVA \cite{1999analysis}. The result shows that \simunote{} and latent vector \textit{z} perform
significantly better than the midi-event tokens in terms of all three criteria (p < 0.005). 

\begin{figure}[htb]
\begin{subfigure}{.5\textwidth}
  \centering
  \includegraphics[width=.88\linewidth]{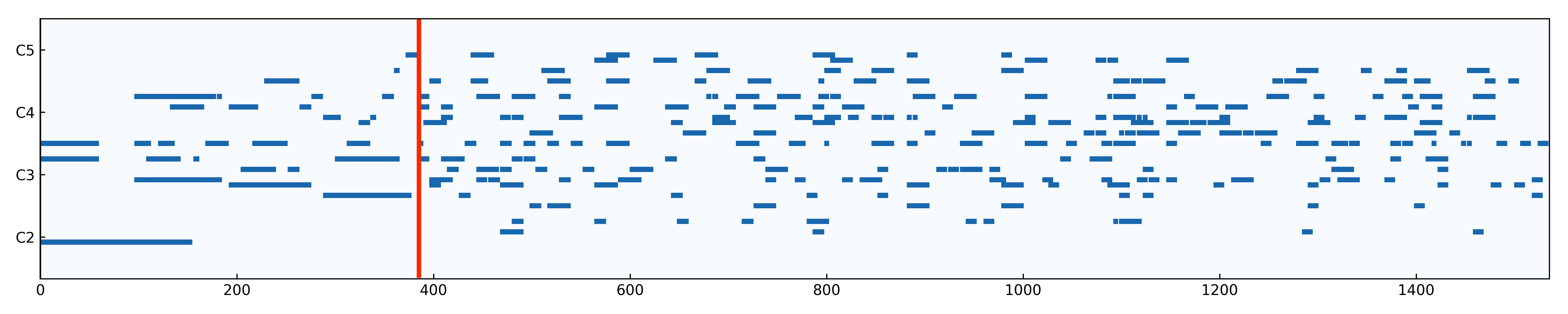}  
  \caption{A sample generated using midi-event tokens.}
  \label{fig:gt}
\end{subfigure}
\begin{subfigure}{.5\textwidth}
  \centering
  \includegraphics[width=.88\linewidth]{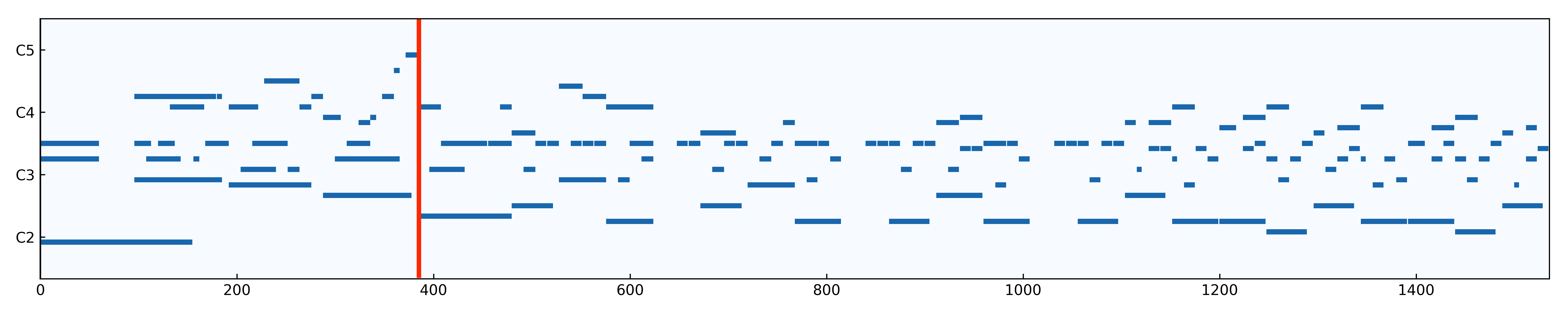}  
  \caption{A sample generated using \simunote{}.}
  \label{fig:simunote}
\end{subfigure}
\begin{subfigure}{.5\textwidth}
  \centering
  \includegraphics[width=.88\linewidth]{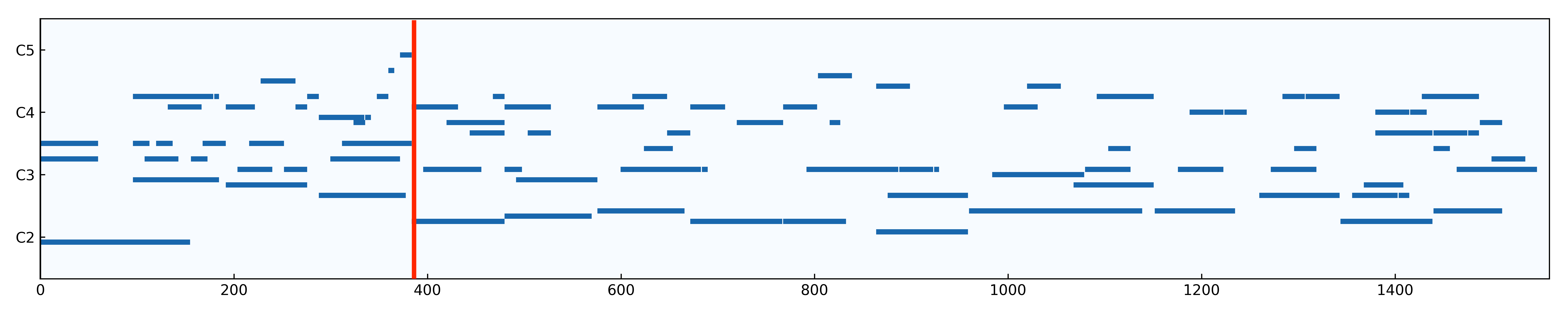}  
  \caption{A sample generated using latent vector \textit{z}.}
  \label{fig:z}
\end{subfigure}
\caption{Long music generations given first 4 measures.}
\label{fig:long generations}
\vspace{-1\baselineskip}
\end{figure}

Besides the aforementioned generation task, we also iteratively feed the generated 4-measure music clips into the model to get longer music compositions. \figref{fig:long generations} shows a comparison of 16-measure generation results using all three representations. The first 4 bars are selected from the test set, and the subsequent 12 bars are generated by the models. Generally speaking, using \simunote{} and latent vector \textit{z} as 
data representations yields more coherent music compositions. Furthermore, we noticed that long generations using the \simunote{} representation tend to repeat previous steps in terms of both chords and rhythms, while those generations using the latent vector \textit{z} usually contain more variations.


\section{Conclusion and Future Work}\label{sec:5}
In conclusion, we proposed PianoTree VAE, a novel representation-learning model tailored for polyphonic music. The key design of the model is to incorporate both the music data structure and model architecture with the sparsity and hierarchical priors. Experiments show that with such inductive biases, PianoTree VAE achieves better reconstruction, interpolation, downstream generation, and strong model interpretability.  
In the future, we plan to extend PianoTree VAE for more general musical structures, such as motif development and multi-part polyphony. 



\newpage
\bibliography{ISMIRtemplate}

\begin{thebibliography}{10}
\providecommand{\url}[1]{#1}
\csname url@samestyle\endcsname
\providecommand{\newblock}{\relax}
\providecommand{\bibinfo}[2]{#2}
\providecommand{\BIBentrySTDinterwordspacing}{\spaceskip=0pt\relax}
\providecommand{\BIBentryALTinterwordstretchfactor}{4}
\providecommand{\BIBentryALTinterwordspacing}{\spaceskip=\fontdimen2\font plus
\BIBentryALTinterwordstretchfactor\fontdimen3\font minus
  \fontdimen4\font\relax}
\providecommand{\BIBforeignlanguage}[2]{{%
\expandafter\ifx\csname l@#1\endcsname\relax
\typeout{** WARNING: IEEEtran.bst: No hyphenation pattern has been}%
\typeout{** loaded for the language `#1'. Using the pattern for}%
\typeout{** the default language instead.}%
\else
\language=\csname l@#1\endcsname
\fi
#2}}
\providecommand{\BIBdecl}{\relax}
\BIBdecl

\bibitem{nlp-vae1}
T.~Zhao, R.~Zhao, and M.~Eskenazi, ``Learning discourse-level diversity for
  neural dialog models using conditional variational autoencoders,''
  \emph{arXiv preprint arXiv:1703.10960}, 2017.

\bibitem{nlp-vae2}
S.~R. Bowman, L.~Vilnis, O.~Vinyals, A.~M. Dai, R.~Jozefowicz, and S.~Bengio,
  ``Generating sentences from a continuous space,'' \emph{arXiv preprint
  arXiv:1511.06349}, 2015.

\bibitem{recurrent-vae}
J.~Chung, K.~Kastner, L.~Dinh, K.~Goel, A.~C. Courville, and Y.~Bengio, ``A
  recurrent latent variable model for sequential data,'' in \emph{Advances in
  neural information processing systems}, 2015, pp. 2980--2988.

\bibitem{musicvae}
A.~Roberts, J.~Engel, C.~Raffel, C.~Hawthorne, and D.~Eck, ``A hierarchical
  latent vector model for learning long-term structure in music,'' \emph{arXiv
  preprint arXiv:1803.05428}, 2018.

\bibitem{ec2vae}
R.~Yang, D.~Wang, Z.~Wang, T.~Chen, J.~Jiang, and G.~Xia, ``Deep music analogy
  via latent representation disentanglement,'' \emph{arXiv preprint
  arXiv:1906.03626}, 2019.

\bibitem{lerchvae}
A.~Pati, A.~Lerch, and G.~Hadjeres, ``Learning to traverse latent spaces for
  musical score inpainting,'' \emph{arXiv preprint arXiv:1907.01164}, 2019.

\bibitem{musegan}
H.-W. Dong, W.-Y. Hsiao, L.-C. Yang, and Y.-H. Yang, ``Musegan: Multi-track
  sequential generative adversarial networks for symbolic music generation and
  accompaniment,'' in \emph{Thirty-Second AAAI Conference on Artificial
  Intelligence}, 2018.

\bibitem{yang2017midinet}
L.-C. Yang, S.-Y. Chou, and Y.-H. Yang, ``Midinet: A convolutional generative
  adversarial network for symbolic-domain music generation,'' \emph{arXiv
  preprint arXiv:1703.10847}, 2017.

\bibitem{mt-vae}
I.~{Simon}, A.~{Roberts}, C.~{Raffel}, J.~{Engel}, C.~{Hawthorne}, and
  D.~{Eck}, ``{Learning a Latent Space of Multitrack Measures},'' \emph{arXiv
  e-prints}, p. arXiv:1806.00195, Jun 2018.

\bibitem{gttm}
F.~Lerdahl and R.~S. Jackendoff, \emph{A generative theory of tonal
  music}.\hskip 1em plus 0.5em minus 0.4em\relax MIT press, 1996.

\bibitem{schenkerian}
J.~Rothgeb, \emph{Introduction to the theory of Heinrich Schenker: the nature
  of the musical work of art}.\hskip 1em plus 0.5em minus 0.4em\relax New York:
  Longman, 1982.

\bibitem{gttmresearch1}
M.~Hamanaka, K.~Hirata, and S.~Tojo, ``Implementing “a generative theory of
  tonal music”,'' \emph{Journal of New Music Research}, vol.~35, no.~4, pp.
  249--277, 2006.

\bibitem{gttmresearch2}
------, ``$\sigma$gttm iii: Learning-based time-span tree generator based on
  pcfg,'' in \emph{International Symposium on Computer Music Multidisciplinary
  Research}.\hskip 1em plus 0.5em minus 0.4em\relax Springer, 2015, pp.
  387--404.

\bibitem{schenresearch1}
S.~W. Smoliar, ``A computer aid for schenkerian analysis,'' in
  \emph{Proceedings of the 1979 annual conference}, 1979, pp. 110--115.

\bibitem{schenresearch2}
A.~Marsden, ``Schenkerian analysis by computer: A proof of concept,''
  \emph{Journal of New Music Research}, vol.~39, no.~3, pp. 269--289, 2010.

\bibitem{deepbach}
G.~Hadjeres, F.~Pachet, and F.~Nielsen, ``Deepbach: a steerable model for bach
  chorales generation,'' in \emph{Proceedings of the 34th International
  Conference on Machine Learning-Volume 70}.\hskip 1em plus 0.5em minus
  0.4em\relax JMLR. org, 2017, pp. 1362--1371.

\bibitem{xiaoice}
H.~Zhu, Q.~Liu, N.~J. Yuan, C.~Qin, J.~Li, K.~Zhang, G.~Zhou, F.~Wei, Y.~Xu,
  and E.~Chen, ``Xiaoice band: A melody and arrangement generation framework
  for pop music,'' in \emph{Proceedings of the 24th ACM SIGKDD International
  Conference on Knowledge Discovery \& Data Mining}.\hskip 1em plus 0.5em minus
  0.4em\relax ACM, 2018, pp. 2837--2846.

\bibitem{huang2017counterpoint}
C.-Z.~A. Huang, T.~Cooijmans, A.~Roberts, A.~Courville, and D.~Eck,
  ``Counterpoint by convolution,'' in \emph{International Society for Music
  Information Retrieval (ISMIR)}, 2017.

\bibitem{music-transformer}
C.-Z.~A. Huang, A.~Vaswani, J.~Uszkoreit, N.~Shazeer, C.~Hawthorne, A.~M. Dai,
  M.~D. Hoffman, and D.~Eck, ``Music transformer: Generating music with
  long-term structure,'' \emph{arXiv preprint arXiv:1809.04281}, 2018.

\bibitem{jambot}
G.~Brunner, Y.~Wang, R.~Wattenhofer, and J.~Wiesendanger, ``Jambot: Music
  theory aware chord based generation of polyphonic music with lstms,'' in
  \emph{2017 IEEE 29th International Conference on Tools with Artificial
  Intelligence (ICTAI)}.\hskip 1em plus 0.5em minus 0.4em\relax IEEE, 2017, pp.
  519--526.

\bibitem{deepj}
H.~H. Mao, T.~Shin, and G.~Cottrell, ``Deepj: Style-specific music
  generation,'' in \emph{2018 IEEE 12th International Conference on Semantic
  Computing (ICSC)}.\hskip 1em plus 0.5em minus 0.4em\relax IEEE, 2018, pp.
  377--382.

\bibitem{rethinking}
E.~S. Koh, S.~Dubnov, and D.~Wright, ``Rethinking recurrent latent variable
  model for music composition,'' in \emph{2018 IEEE 20th International Workshop
  on Multimedia Signal Processing (MMSP)}.\hskip 1em plus 0.5em minus
  0.4em\relax IEEE, 2018, pp. 1--6.

\bibitem{boulangerg}
N.~Boulanger-Lewandowski, Y.~Bengio, and P.~Vincent, ``Modeling temporal
  dependencies in high-dimensional sequences: Application to polyphonic music
  generation and transcription,'' \emph{arXiv preprint arXiv:1206.6392}, 2012.

\bibitem{perf-rnn}
I.~Simon and S.~Oore, ``Performance rnn: Generating music with expressive
  timing and dynamics,'' \url{https://magenta.tensorflow.org/performance-rnn},
  2017.

\bibitem{lakhnes}
C.~Donahue, H.~H. Mao, Y.~E. Li, G.~W. Cottrell, and J.~McAuley, ``Lakhnes:
  Improving multi-instrumental music generation with cross-domain
  pre-training,'' \emph{arXiv preprint arXiv:1907.04868}, 2019.

\bibitem{poptfm}
Y.-S. Huang and Y.-H. Yang, ``Pop music transformer: Generating music with
  rhythm and harmony,'' \emph{arXiv preprint arXiv:2002.00212}, 2020.

\bibitem{crnngan}
O.~Mogren, ``C-rnn-gan: Continuous recurrent neural networks with adversarial
  training,'' \emph{arXiv preprint arXiv:1611.09904}, 2016.

\bibitem{tfm-nade}
C.~Hawthorne, A.~Huang, D.~Ippolito, and D.~Eck, ``Transformer-nade for piano
  performances.''

\bibitem{gnn}
F.~Scarselli, M.~Gori, A.~C. Tsoi, M.~Hagenbuchner, and G.~Monfardini, ``The
  graph neural network model,'' \emph{IEEE Transactions on Neural Networks},
  vol.~20, no.~1, pp. 61--80, 2008.

\bibitem{musicgnn}
D.~Jeong, T.~Kwon, Y.~Kim, and J.~Nam, ``Graph neural network for music score
  data and modeling expressive piano performance,'' in \emph{International
  Conference on Machine Learning}, 2019, pp. 3060--3070.

\bibitem{syntax1}
C.~Dyer, A.~Kuncoro, M.~Ballesteros, and N.~A. Smith, ``Recurrent neural
  network grammars,'' \emph{arXiv preprint arXiv:1602.07776}, 2016.

\bibitem{syntax2}
K.~S. Tai, R.~Socher, and C.~D. Manning, ``Improved semantic representations
  from tree-structured long short-term memory networks,'' \emph{arXiv preprint
  arXiv:1503.00075}, 2015.

\bibitem{gru}
K.~Cho, B.~Van~Merri{\"e}nboer, C.~Gulcehre, D.~Bahdanau, F.~Bougares,
  H.~Schwenk, and Y.~Bengio, ``Learning phrase representations using rnn
  encoder-decoder for statistical machine translation,'' \emph{arXiv preprint
  arXiv:1406.1078}, 2014.

\bibitem{vae}
D.~P. Kingma and M.~Welling, ``Auto-encoding variational bayes,'' \emph{arXiv
  preprint arXiv:1312.6114}, 2013.

\bibitem{betavae}
I.~Higgins, L.~Matthey, A.~Pal, C.~Burgess, X.~Glorot, M.~Botvinick,
  S.~Mohamed, and A.~Lerchner, ``beta-vae: Learning basic visual concepts with
  a constrained variational framework,'' 2016.

\bibitem{pop909}
Z.~Wang, K.~Chen, J.~Jiang, Y.~Zhang, M.~Xu, S.~Dai, X.~Gu, and G.~Xia,
  ``Pop909: A pop-song dataset for music arrangement generation,'' in
  \emph{Proceedings of 21st International Conference on Music Information
  Retrieval ({ISMIR}), virtual conference}, 2020.

\bibitem{adam}
D.~P. Kingma and J.~Ba, ``Adam: A method for stochastic optimization,''
  \emph{arXiv preprint arXiv:1412.6980}, 2014.

\bibitem{teacherforcing}
N.~B. Toomarian and J.~Barhen, ``Learning a trajectory using adjoint functions
  and teacher forcing,'' \emph{Neural networks}, vol.~5, no.~3, pp. 473--484,
  1992.

\bibitem{metric}
C.~Hawthorne, E.~Elsen, J.~Song, A.~Roberts, I.~Simon, C.~Raffel, J.~Engel,
  S.~Oore, and D.~Eck, ``Onsets and frames: Dual-objective piano
  transcription,'' \emph{arXiv preprint arXiv:1710.11153}, 2017.

\bibitem{mir_eval}
C.~Raffel, B.~McFee, E.~J. Humphrey, J.~Salamon, O.~Nieto, D.~Liang, D.~P.
  Ellis, and C.~C. Raffel, ``mir\_eval: A transparent implementation of common
  mir metrics,'' in \emph{In Proceedings of the 15th International Society for
  Music Information Retrieval Conference, ISMIR}.\hskip 1em plus 0.5em minus
  0.4em\relax Citeseer, 2014.

\bibitem{our-nime-vae}
R.~Yang, T.~Chen, Y.~Zhang, and G.~Xia, ``Inspecting and interacting with
  meaningful music representations using vae,'' \emph{arXiv preprint
  arXiv:1904.08842}, 2019.

\bibitem{slerp}
A.~Watt and M.~Watt, ``Advanced animatidn and bendering technidues,'' 1992.

\bibitem{ke}
K.~Chen, G.~Xia, and S.~Dubnov, ``Continuous melody generation via disentangled
  short-term representations and structural conditions,'' \emph{2020 IEEE 14th
  International Conference on Semantic Computing (ICSC)}, pp. 128--135, 2020.

\bibitem{tfm}
A.~Vaswani, N.~Shazeer, N.~Parmar, J.~Uszkoreit, L.~Jones, A.~N. Gomez,
  L.~Kaiser, and I.~Polosukhin, ``Attention is all you need,'' \emph{ArXiv},
  vol. abs/1706.03762, 2017.

\bibitem{1999analysis}
H.~Scheffe, \emph{The analysis of variance}.\hskip 1em plus 0.5em minus
  0.4em\relax John Wiley \& Sons, 1999, vol.~72.

\end{thebibliography}

%
%
%
%

\end{document}